\documentclass[12pt]{article}
\pdfoutput=1
\usepackage{subfigure}
\usepackage{amssymb,amsmath}
\usepackage{graphicx}
\usepackage{color}
\usepackage[colorlinks=true
,urlcolor=blue
,citecolor=blue
,linkcolor=blue
,pagecolor=blue
,linktocpage=true
,pdfproducer=medialab
]{hyperref}
\usepackage[a4paper,width=15.2cm]{geometry}
\makeatletter \renewcommand{\@dotsep}{10000} \makeatother

\newcommand{\beq}{\begin{equation}}
\newcommand{\eeq}{\end{equation}}
\newcommand{\bea}{\begin{eqnarray}}
\newcommand{\eea}{\end{eqnarray}}

\begin{document}
%Remove date before submitting to arXiv
%\date{\today}

\begin{flushright}
 CETUP2013-009
\end{flushright}

\vspace{1cm}
\begin{center}

 {\Large\bf  Sparticle Spectroscopy from SO(10) GUT with a Unified Higgs Sector
 } \vspace{1cm}

{\large  M. Adeel Ajaib$^{a,}$\footnote{E-mail: adeel@udel.edu}, Ilia Gogoladze$^{a,}$\footnote{E-mail:
ilia@bartol.udel.edu\\ \hspace*{0.5cm} On  leave of absence from:
Andronikashvili Institute of Physics, 0177 Tbilisi, Georgia.} and   Qaisar Shafi$^{a,}$\footnote{E-mail:
shafi@bartol.udel.edu} } \vspace{.5cm}

{ \it
$^a$Bartol Research Institute, Department of Physics and Astronomy, \\
University of Delaware, Newark, DE 19716, USA
} \vspace{.5cm}

\vspace{1.5cm}
 {\bf Abstract}
\end{center}

We study the low energy implications, especially the particle spectroscopy, of SO(10) grand unification in which the SO(10) symmetry is broken to the Standard Model gauge group with a single pair of $(144+\overline{144})$ dimensional Higgs multiplet (unified Higgs sector).
In this  class of models, the asymptotic relation $Y_b\approx Y_\tau\approx Y_t/6$ among the third generation quark and lepton Yukawa couplings  can be derived.  This relation leads to the prediction $\tan\beta \approx 14$, where $\tan\beta$ is the well known MSSM parameter. We find that this type of Yukawa coupling unification (YU) is realized only by employing non-universal soft supersymmetry breaking terms, dictated by SO(10) symmetry, for the gauginos. A 125 GeV Higgs boson mass is also found to be consistent with YU at the $\sim 5\%$ level.
Without imposing a constraint on the relic abundance of dark matter in these models, the squark and  slepton masses, with the exception of the stop,  exceed 2 TeV and the gluino is heavier than 1 TeV.
We show that the neutralino in this model is an acceptable dark matter candidate through the neutralino-stop coannihilation scenario, with the stop quark being relatively light ($\gtrsim 500$ GeV).

\newpage

%%%%%%%%%%%%%%%%%%%%%%%%%%%%%%%%%%%%%%%%%%%%%%%%%%%%%%%%%%%%
\renewcommand{\thefootnote}{\arabic{footnote}}
\setcounter{footnote}{0}

%%%%%%%%%%%%%%%%%%%%%%%%%%%%%%%%%%%%%%%%%%%%%%%%%%%%%%%%%%%%%

%\baselineskip 36pt
% Main body
%%%%%%%%%%%%%%%%%%%%%%%%%%
%\baselineskip 18pt
%%%%%%%%%%%%%%%%%%%%%%%%%%

%%%%%%%%%%%%%%%%%%%%%%%%%%%%%%%%%%%%%%%%%%%%%%%%%%%%%%%%%%%%
\renewcommand{\thefootnote}{\arabic{footnote}}
\setcounter{footnote}{0}

%%%%%%%%%%%%%%%%%%%%%%%%%%%%%%%%%%%%%%%%%%%%%%%%%%%%%%%%%%%%%

%\baselineskip 36pt
% Main body
%%%%%%%%%%%%%%%%%%%%%%%%%%
%\baselineskip 18pt
%%%%%%%%%%%%%%%%%%%%%%%%%%

\section{Introduction}

{An SO(10)} gauge symmetry provides {an elegant} framework for unifying  the strong and electroweak interactions. A single generation of quarks and leptons {including a right handed neutrino, nicely fits} in an irreducible 16 dimensional representation  \cite{georgi}. The right handed neutrino ($\nu_R$) helps generate {the observed} light neutrino masses via the
see-saw mechanism \cite{seesawI}. It can also naturally account for the observed baryon asymmetry of the universe
via leptogenesis \cite{Fukugita:1986hr}.
Another virtue of SO(10) is that, in principle, the two MSSM Higgs doublets can be accommodated in a single ten dimensional representation ($10_{\rm H}$), which then yields the following Yukawa couplings
\begin{align}
Y_{ij}\ 16_i\, 16_j\, 10_{\rm H}.
\label{10-yukawa}
\end{align}
Here $i,j=1, 2, 3$ stand for family indices and the SO(10) indices have been omitted for simplicity.
{Considering only} the third generation quarks and  leptons, the interaction  in Eq.(\ref{10-yukawa}) yields  the following Yukawa coupling unification (YU) condition  at $M_{GUT}$:
\begin{align}
Y_t = Y_b = Y_{\tau} = Y_{\nu_{\tau}}. \label{f1}
\end{align}
{where $Y_{\nu_{\tau}}$ denotes the tau neutrino Dirac coupling}. Consequently, large $\tan\beta\sim 50$ is predicted  \cite{big-422} in order to get compatibility with {experimental observations}.

 It is interesting to note that in the gravity mediation SUSY breaking scenario \cite{Chamseddine:1982jx}, $t$-$b$-$\tau$ {YU condition leads to} LHC testable sparticle spectrum \cite{bigger-422} and {it even `predicts'} a 125 GeV light CP-even Higgs boson mass \cite{Gogoladze:2011aa}.

One {potential} drawback of SO(10) grand {unification}  is the
lack of  a unique minimal model due to the {various} possibilities {available in} the Higgs  sector {for breaking} $SO(10)$ {to} $SU(3)_C\times U(1)_{em}$. Typically, one needs  a $16+\overline{16}$
or a $126+\overline{126}$  Higgs representation to reduce the rank of the group {from five to four together} with either  a 45 or 210-dimensional representation  for breaking the symmetry
down to $SU(3)_C \times SU(2)_L \times U(1)_Y$.
 Furthermore, one needs a $10$-dimensional
{Higgs multiplet} to break the electroweak symmetry.
These requirements imply that {in principle,} two {distinct superheavy} mass scales are involved in the breaking of SO(10), one {associated with the reduction of} the rank, and the other {for breaking} the symmetry all the way down to the Standard Model (SM).
 In order to maintain  gauge and Yukawa coupling unification,  one needs to assume that the various vacuum expectation values (VEVs) are of the same order {of magnitude}.  This requires {suitable}  relations among the  parameters in the superpotential which  {may not appear very} natural.

Recently a new class of  SO(10) models  was presented \cite{Babu:2005gx,Babu:2006rp} where the {SO(10)} symmetry breaking down to the SM gauge group involves just  a single pair of ($144+\overline{144}$)-dimensional vector-spinor  Higgs multiplet. It was also shown that {this pair of multiplets} can contain a pair of light Higgs doublets, necessary for breaking the electroweak symmetry. In order to solve the doublet-triplet splitting problem in this class of models, a ($560+\overline{560}$)-dimensional vector-spinor representation was introduced
instead of $144+\overline{144}$ \cite{Babu:2011tw}.  In this case the  doublet-triplet splitting problem was solved via the missing partner mechanism.

In this paper we study the low {energy} spectrum  of supersymmetric SO(10) GUT with $144+\overline{144}$ dimensional Higgs multiplet. The Yukawa coupling for third generation quarks and leptons {is given by}
\begin{align}
Y_{i}\ \frac{16_i\, 16_i\, 144_{\rm H}\, 144_{\rm H}}{M_*},
\label{eq144-1}
\end{align}
where $M_*$ is a super heavy mass.
It was shown in ref.  \cite{Babu:2005gx} that there {corresponds a}  parameter space in this class of model where the following relation among third generation  quark and lepton Yukawa couplings is obtained:
\begin{align}
 Y_b \approx Y_{\tau} \approx \frac{Y_t }{6}. \label{eq144-2}
\end{align}
In order to be compatible with {observations}, the theory predicts an intermediate value $\sim 10$ for  $\tan\beta$.

In this paper we seek the low scale sparticle spectrum which {is consistent with} the asymptotic relation presented in Eq. (\ref{eq144-2}). We find that this requires non-universal gaugino masses at $M_{GUT}$ which, {as previously discussed}, can be incorporated in the SO(10) framework.

The outline for the rest of the paper is as follows.
In Section \ref{theory} we present the parameter space that we randomly scan and describe how the MSSM gaugino mass relations can be obtained at $M_{\rm GUT}$. In Section \ref{constraintsSection} we summarize the scanning procedure and the experimental constraints applied in our analysis. The results are  presented in Section \ref{results}. The table in this section lists some benchmark points which can be tested at the LHC. Our conclusions are summarized in Section \ref{conclusions}.

\section{Fundamental Parameter Space \label{theory}}

We first comment on our results when the asymptotic relation among the Yukawa couplings presented in Eq.(\ref{eq144-2}) {is applied assuming universal} gaugino masses at $M_{GUT}$. We find that in this case the Yukawa coupling unification {is not} better than the $30\%$ level, {regardless of} whether universal or non-universal  Higgs soft supersymmetry (SUSY) breaking mass terms are imposed.
Based on the {experience} (see for instance ref. \cite{Gogoladze:2011aa}) that non-universal gaugino masses help achieve
conventional  Yukawa unification ($Y_t=Y_b=Y_{\tau}$), we employ the same non-universal gaugino mass condition in this analysis as well. We will show in section \ref{results} that non-universal gaugino masses also leads to unification of Yukawa couplings {according to} Eq.(\ref{eq144-2}).

 It has been pointed out \cite{Martin:2009ad} that non-universal MSSM gaugino masses at $ M_{\rm GUT} $ can arise from non-singlet {SUSY breaking} F-terms, compatible with the underlying GUT symmetry. The SSB
gaugino masses in supergravity  \cite{Chamseddine:1982jx} can arise  from the following
dimension five operator:
\begin{align}
 -\frac{F^{ab}}{2 M_{\rm
P}} \lambda^a \lambda^b + {\rm c.c.}
\end{align}
 Here $\lambda^a$ is the two-component gaugino field, $ F^{ab} $ denotes the F-component of the field which breaks SUSY, and the indices $a,b$ run over
the adjoint representation {of the gauge group}. The resulting gaugino
mass matrix is $\langle F^{ab} \rangle/M_{\rm P}$, where the
SUSY breaking  parameter $\langle F^{ab} \rangle$
transforms as a singlet under the MSSM gauge group $SU(3)_{c}
\times SU(2)_L \times U(1)_Y$. The $F^{ab}$ fields belong to an
irreducible representation in the symmetric part of the direct product of the
adjoint representation of the unified group.

In SO(10), for example,
\begin{align}
({ 45} \times { 45} )_S = { 1} + { 54} + { 210} +
{ 770}.
\end{align}
If  $F$  transforms as a 54 or 210 dimensional
representation of SO(10) \cite{Martin:2009ad}, one obtains the following relation
among the MSSM gaugino masses at $ M_{\rm GUT} $ :
\begin{align}
M_3: M_2:M_1= 2:-3:-1 ,
\label{gaugino10}
\end{align}
where $M_1, M_2, M_3$ denote the gaugino masses of $U(1)$, $SU(2)_L$ and $SU(3)_c$
respectively.

Notice that in  general, if $F^{ab}$ transforms non trivially under SO(10), the SSB terms such as the trilinear couplings and scalar mass terms are not necessarily universal at $M_{GUT}$. However, we can assume, consistent with SO(10) gauge symmetry, that the coefficients associated with terms that violate the SO(10)-invariant form are suitably small, except for the gaugino terms in Eq.(\ref{gaugino10}). We also assume that D-term contributions to the SSB terms are much smaller compared with contributions from fields with non-zero auxiliary F-terms.

Employing the boundary condition from Eq.(\ref{gaugino10}), we define the MSSM gaugino masses at $ M_{\rm GUT} $ in terms of the mass parameter $M_{1/2}$ :
\begin{align}
M_1= M_{1/2},~~
M_2= 3M_{1/2} ~~\rm{and}~~
M_3= - 2 M_{1/2}.
 \label{gaugino11}
\end{align}

We have performed random scans for the following parameter range:
\begin{align}
0\leq  m_{16}  \leq 20\, \rm{TeV}, ~~~
0\leq   m_{144} \leq 20\, \rm{TeV}, \nonumber \\
0 \leq M_{1/2}  \leq 5 \, \rm{TeV}, ~~~
-3\leq A_{0}/m_{16} \leq 3, \nonumber \\
2\leq \tan\beta \leq 60.
 \label{parameterRange}
\end{align}
 Here $ m_{16} $ is the universal SSB mass for MSSM sfermions, $ m_{144} $ is the universal SSB mass term for up and down MSSM Higgs masses, $ M_{1/2} $ is the gaugino mass parameter, $ \tan\beta $ is the ratio of the vacuum expectation values (VEVs) of the two MSSM Higgs doublets, and $ A_{0} $ is the universal SSB trilinear scalar interaction (with corresponding Yukawa coupling factored out).  We use {the} central value $m_t = 173.1\, {\rm GeV}$ and 1$\sigma$ deviation    ($m_t = 174.2\, {\rm GeV}$) for top quark  in our analysis   \cite{:1900yx}. {A +1$\sigma$ increase in $m_t$ slightly raises the Higgs mass which is desirable in our analysis}. Our results however are not
too sensitive to one or two sigma variation in the value of $m_t$  \cite{Gogoladze:2011db}.
We use $m_b(m_Z)=2.83$ GeV which is hard-coded into Isajet.

%%%%%%%%%%%%%%%%%%%%%%%%%%%%%%%%%%%%%%%%%%%%%%%%%%%%%%%%%%%%%%%%%%%%%%%%%%%%%%%%%%%%%%%%%%%%%%%%%%%%%%%%%

\section{Phenomenological Constraints and Scanning Procedure\label{constraintsSection}}

We employ the ISAJET~7.84 package~\cite{ISAJET}  to perform random
scans over the fundamental parameter space. In this package, the weak scale values of gauge and third generation Yukawa
couplings are evolved to $M_{\rm GUT}$ via the MSSM renormalization
group equations (RGEs) in the $\overline{DR}$ regularization scheme.
We do not strictly enforce the unification condition $g_3=g_1=g_2$ at $M_{\rm
GUT}$, since a few percent deviation from unification can be
assigned to unknown GUT-scale threshold
corrections~\cite{Hisano:1992jj}.
The deviation between $g_1=g_2$ and $g_3$ at $M_{\rm GUT}$ is no
worse than $3-4\%$.
For simplicity  we do not include the Dirac neutrino Yukawa coupling
in the RGEs, whose contribution is expected to be small.

The various boundary conditions are imposed at
$M_{\rm GUT}$ and all the SSB
parameters, along with the gauge and Yukawa couplings, are evolved
back to the weak scale $M_{\rm Z}$.
In the evaluation of Yukawa couplings the SUSY threshold
corrections~\cite{Pierce:1996zz} are taken into account at the
common scale $M_{\rm SUSY}= \sqrt{m_{{\tilde t}_L}m_{{\tilde t}_R}}$,
where $m_{{\tilde t}_L}$ and $m_{{\tilde t}_R}$
are the third generation left and right handed stop quark masses.
 The entire
parameter set is iteratively run between $M_{\rm Z}$ and $M_{\rm
GUT}$ using the full 2-loop RGEs until a stable solution is
obtained. To better account for leading-log corrections, one-loop
step-beta functions are adopted for gauge and Yukawa couplings, and
the SSB parameters $m_i$ are extracted from RGEs at {their appropriate} scales
$m_i=m_i(m_i)$. The RGE-improved 1-loop effective potential is
minimized at $M_{\rm SUSY}$, which effectively
accounts for the leading 2-loop corrections. Full 1-loop radiative
corrections are incorporated for all sparticle masses.

An important constraint comes from limits on the cosmological abundance of stable charged
particles  \cite{Nakamura:2010zzi}. This excludes regions in the parameter space
where  charged SUSY particles  become
the lightest supersymmetric particle (LSP). We accept only those
solutions for which one of the neutralinos is the LSP and saturates
the WMAP  bound on relic dark matter abundance.

An approximate error  of around  $2$ GeV in the estimate of the
Higgs mass in Isajet  largely arises from theoretical  uncertainties  in the calculation of the minimum of the scalar potential,  and
to a lesser extent from experimental uncertainties in the values
for $m_t$ and $\alpha_s$.

Micromegas 2.4 \cite{Belanger:2008sj} is interfaced with Isajet to calculate the relic density and branching ratios $BR(B_s \rightarrow \mu^+ \mu^-)$ and $BR(b \rightarrow s \gamma)$. We implement the following random scanning procedure: A uniform and logarithmic distribution of random points is first generated in the parameter space given in Eq. (\ref{parameterRange}).
The function RNORMX \cite{Leva} is then employed
to generate a Gaussian distribution around each point in the parameter space.  The data points
collected all satisfy
the requirement of radiative electroweak symmetry breaking  (REWSB),
with the neutralino in each case being the LSP. After collecting the data, we impose
the mass bounds on all the particles \cite{Nakamura:2010zzi} and use the
IsaTools package~\cite{Baer:2002fv}
to implement the various phenomenological constraints. We successively apply the following experimental constraints on the data that
we acquire from SuSpect and Isajet:

\begin{table}[h]\centering
\begin{tabular}{rlc}
$ 0.8 \times 10^{-9} \leq BR(B_s \rightarrow \mu^+ \mu^-) $&$ \leq\, 6.2 \times 10^{-9} \;
 (2\sigma)$        &   \cite{:2007kv}      \\
$2.99 \times 10^{-4} \leq BR(b \rightarrow s \gamma) $&$ \leq\, 3.87 \times 10^{-4} \;
 (2\sigma)$ &   \cite{Barberio:2008fa}  \\
$0.15 \leq \frac{BR(B_u\rightarrow
\tau \nu_{\tau})_{\rm MSSM}}{BR(B_u\rightarrow \tau \nu_{\tau})_{\rm SM}}$&$ \leq\, 2.41 \;
(3\sigma)$ &   \cite{Barberio:2008fa}  \\
 $ 0 \leq \Delta(g-2)_{\mu}/2 $ & $ \leq 55.6 \times 10^{-10} $ & \cite{Bennett:2006fi}
\end{tabular}\label{table}
\end{table}

In order to quantify Yukawa coupling unification, we define the quantity $R^{\prime}_{tb\tau}$ as,
\begin{align}
R^{\prime}_{tb\tau}=\frac{ {\rm max}(y_t/6,y_b,y_{\tau})} { {\rm min} (y_t/6,y_b,y_{\tau})}.
\end{align}
%

%%%%%%%%%%%%%%%%%%%%%%%%%%%%%%%%%%%%%%%%%%%%%%%%%%%%%%%%%%%%%%%%%%%%%%%%%%%%%%

%%%%%%%%%%%%%%%%%%%%%%%%%%%%%%%%%%%%%%%%%%%%%%%%%%%%%%%%%%%
%trim=l b r t
%\newpage
\begin{figure}[]
%\newpage
%\vspace{-1cm}
\centering
{\label{fig:4-b}{\includegraphics[scale=0.4]{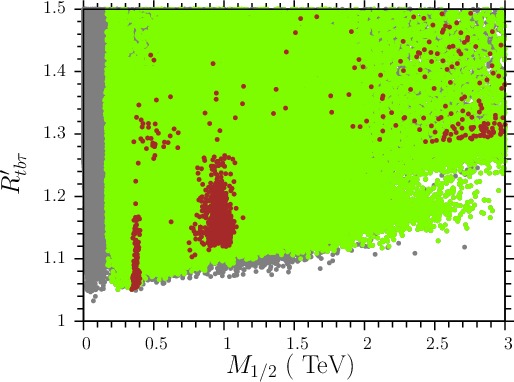}}}
{\label{fig:4-b}{\includegraphics[scale=0.4]{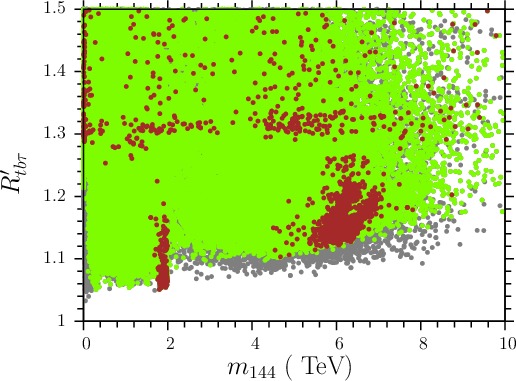}}}
{\label{fig:4-b}{\includegraphics[scale=0.4]{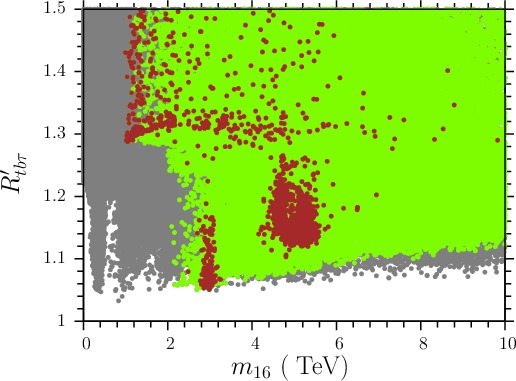}}}
{\label{fig:4-b}{\includegraphics[scale=0.4]{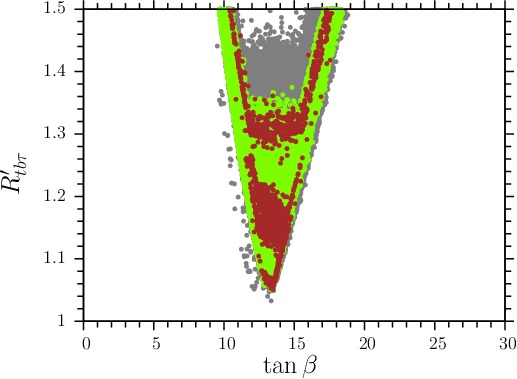}}}
\caption{Plots in the $R^{\prime}_{tb\tau} - M_{1/2}$, $R^{\prime}_{tb\tau} - m_{144}$, $R^{\prime}_{tb\tau} - m_{16}$ and $R^{\prime}_{tb\tau} - \tan\beta$ planes. The data points shown are collected using Isajet 7.84. Gray points are consistent with REWSB and LSP neutralino. Green points form a subset of the gray and satisfy sparticle mass and B-physics constraints. The green points also satisfy the Higgs mass bound $123 {\rm \ GeV} \le m_h \le 127 \rm \ GeV$.
  Moreover, we require that the green points do no worse than the SM in terms of $ (g-2)_{\mu} $.
 Brown points form a subset of the green points and satisfy $\Omega h^2 \le 10$. }
\label{fig:nugm1}
\end{figure}

%%%%%%%%%%%%%%%%%%%%%%%%%%%%%%%%%%%%%%%%%%%%%%\vspace*{2mm}

%%%%%%%%%%%%%%%%%%%%%%%%%%%%%%%%%%%%%%%%%%%%%%\vspace*{2mm}
%
\begin{figure}[h!]
%\newpage
%\vspace{-1cm}
\centering
{\label{fig:4-b}{\includegraphics[scale=0.39]{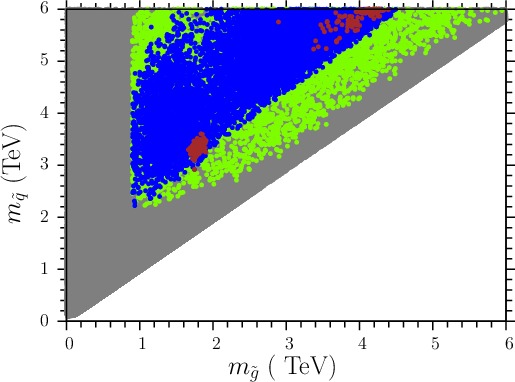}}}
{\label{fig:4-b}{\includegraphics[scale=0.39]{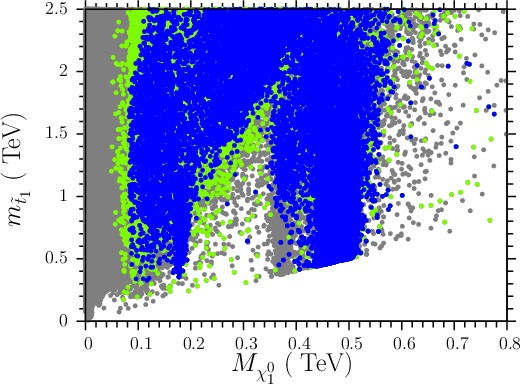}}}\\
\vspace{0.3cm}
{\label{fig:4-b}{\includegraphics[scale=0.39]{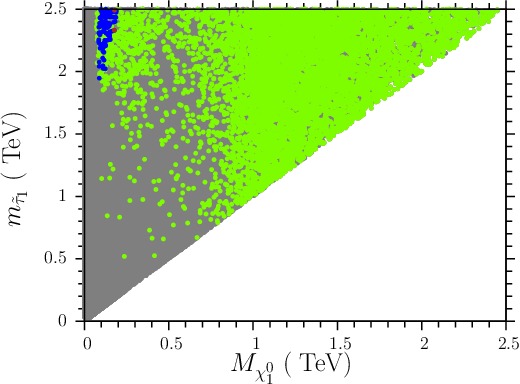}}}
{\label{fig:4-b}{\includegraphics[scale=0.39]{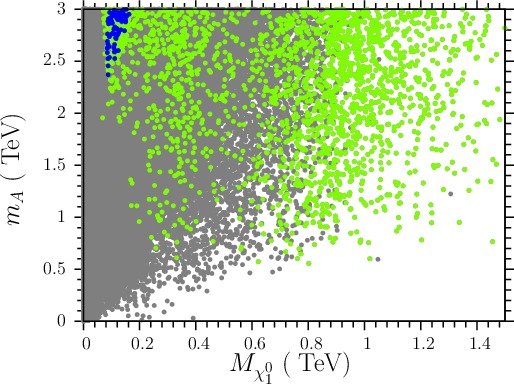}}}
\caption{Plots in the $R^{\prime}_{tb\tau} - M_{\chi_1^0}$, $m_{\tilde{\tau}_1} - M_{\chi_1^0}$, $m_A -M_{\chi_1^0}$ and $m_{\tilde{t}_1} - M_{\chi_1^0}$ planes.  Color coding is the same as in Figure \ref{fig:nugm1}. In addition   blue points are subset of the green and satisfy $R^{\prime}_{tb\tau} < 1.2$. Brown points form a subset of the blue points and satisfy $\Omega h^2 \le 10$. }
\label{fig:nugm3}
\end{figure}

\section{Sparticle Spectroscopy and the Higgs mass  \label{results}}

 Figure \ref{fig:nugm1} shows our results in the $R^{\prime}_{tb\tau} - M_{1/2}$, $R^{\prime}_{tb\tau} - m_{10}$, $R^{\prime}_{tb\tau} - m_{16}$ and $R^{\prime}_{tb\tau} - \tan\beta$ planes. Gray points in the figure are consistent with REWSB and LSP neutralino. Green points form a subset of the gray and satisfy sparticle mass and B-physics constraints described in section \ref{constraintsSection}. The green points also satisfy the Higgs mass bound $123 {\rm \ GeV} \le m_h \le 127 \rm \ GeV$.  Brown points form a subset of the green points and satisfy $\Omega h^2 \le 10$. We chose to concentrate on  $\Omega h^2 \le 10$ because in this model neutralino is mostly a bino like particle and it is heavier than a 100 GeV. In this case, without any additional contribution, $\Omega h^2$ can be around $O(10^2)$ or even $O(10^3)$ \cite{Baer:2006te}.    {So, $\Omega h^2 \le 10$ already indicates that there is some additional mechanism which significantly reduces the relic abundance  close to the desired value with some fine tuning in the SSB parameter space}.

Figure  \ref{fig:nugm1} shows that our analysis does not yield better than $\sim 5\%$ YU consistent with the constraints described in section \ref{constraintsSection}. {The prediction for YU essentially remains the same if we require the relic density to be small, $\Omega h^2 \le 10$}.  We also observe that requiring good YU leads to narrow ranges of the fundamental parameters in the model. The gaugino mass parameter ($M_{1/2}$)  consistent with good YU is $\sim 200$ GeV whereas the Higgs mass parameter ($m_{144}$) lies in the range $2-4 \rm \ TeV$.
Similarly, good YU prefers the GUT scale scalar mass {parameter} $m_{16} \sim 2$ TeV and $\tan\beta \sim 14$. Note that {this value} for $\tan\beta$ is notably different from the {prediction} $\tan\beta \sim 47$ in refs. \cite{Gogoladze:2011aa}, which {studied} the same model {but }with the condition $Y_t=Y_b=Y_{\tau}$ at $M_{GUT}$.

Note also that requiring the {neutralino} relic abundance of the neutralino to satisfy $\Omega h^2 \le 10$ affects the above mentioned predictions. While the preferred value for $\tan\beta$ essentially remains the same, the smallest values of the parameters   $M_{1/2}$, $m_{144}$ and $m_{16}$ consistent with {YU $\sim 5\%$ are pushed to higher values, namely, $M_{1/2} \sim 300 \rm \ GeV$, $m_{144} \sim 1.6 \rm \ TeV$ and $m_{16} \sim 2.8 \rm \ TeV$.}

%%%%%%%%%%%%%%%%%%%%%%%%%%%%%%%%%%%%%%%%%%%%%%%%%%%%%%%%%%%

\begin{figure}[]
\begin{center}
\includegraphics[scale=0.38]{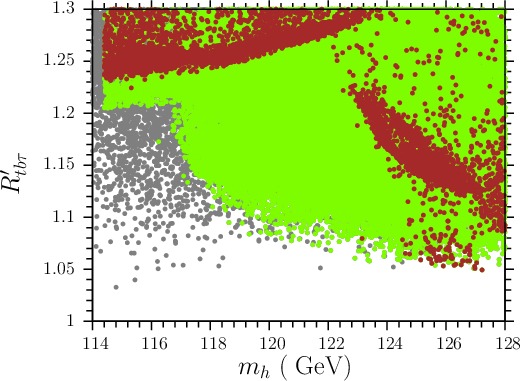}
\includegraphics[scale=0.38]{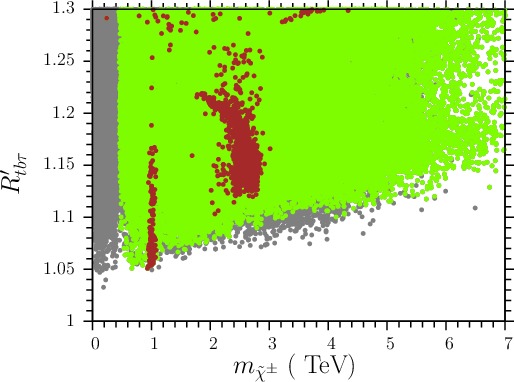}
%\vspace{0.3cm}
%\includegraphics[scale=0.4]{plot-ratio-mchi-nugm.jpg}
%\includegraphics[scale=0.4]{plot-ratio-mcha-nugm.jpg}
\end{center}
\caption{Plots in the $R^{\prime}_{tb\tau} - m_h$ and  $R^{\prime}_{tb\tau} - m_{\tilde{\chi}^{\pm}}$  planes. The color coding is the same as in Figure \ref{fig:nugm1}.}
\label{fig-144-2}
\end{figure}

%%%%%%%%%%%%%%%%%%%%%%%%%%%%%%%%%%%%%%%%%%%%%%%%%%%%%%%%%%%%%%%%%%%%

Figure \ref{fig:nugm3} shows plots in the $m_{\tilde{q}}-m_{\tilde{g}}$, $m_{\tilde{\tau}_1} - M_{\chi_1^0}$, $m_A -M_{\chi_1^0}$ and $m_{\tilde{t}_1} - M_{\chi_1^0}$ planes. The green points have the same definition as in Figure \ref{fig:nugm1}. The blue points are a subset of the green and satisfy $R^{\prime}_{tb\tau} < 1.2$.
Brown points form a subset of the blue points and satisfy $\Omega h^2 \le 10$.
The $m_{\tilde{q}}-m_{\tilde{g}}$ plane shows that 20\%  or better YU  predicts the first and second generation squark masses to be $\gtrsim 2 \rm \ TeV$.
The  $m_{\tilde{t}_1} - M_{\chi_1^0}$ shows that neutralino-stop coannihilation is consistent with good YU. {We can see from the $m_{\tilde{q}}-m_{\tilde{g}}$ plane that for $\Omega h^2 \le 10$ (brown points)  the first two generation squarks have masses $\gtrsim 3 \rm \ TeV$ and gluinos are {heavier than} $1.5 \rm \ TeV$ or so.}
 We can conclude, based on the location of blue points in the $m_{\tilde{\tau}_1} - M_{\chi_1^0}$ plane,
that neutralino-stau coannihilation is impossible to realize in this model. From the $m_A -M_{\chi_1^0}$ plot, we {observe} that good YU is not consistent with the $M_A$ resonance solution for neutralino dark matter.

%%%%%%%%%%%%%%%%%%%%%%%%%%%%%%%%%%%%%%%%%%%%%%\vspace*{2mm}
\begin{table}[h!p]
%\vspace{1.5cm}
\centering
\begin{tabular}{|p{3cm}|p{3cm}p{3cm}p{3cm}|}
\hline
\hline
                 	&	 Point 1	&	 Point 2 	&	 Point 3 	\\
%\times 10^{							
\hline							
%\times 10^{- \times 10^{ \times 10^{							
							
$m_{10} $         	&$	5936.55	$&$	1479.57	$&$	1086.04	$\\
$m_{16} $         	&$	4958.8	$&$	2922.23	$&$	2741.97	$\\
$M_1$         	&$	928.11	$&$	406.53	$&$	315.61	$\\
$M_2$         	&$	2784.33	$&$	1219.58	$&$	946.83	$\\
$M_3$         	&$	-1856.22	$&$	-813.06	$&$	-631.22	$\\
$A_0/m_{16}$         	&$	-2.88	$&$	-2.98	$&$	-2.99	$\\
$\tan\beta$      	&$	14.06	$&$	13.42	$&$	13.44	$\\
%$m_t$            	&$	173.1	$&$	173.1	$&$	173.1	$\\
\hline		  		  		  	
$\mu$            	&$	3652	$&$	3448	$&$	3315	$\\

\hline		  		  		  	
$m_h$            	&$	126	$&$	124	$&$	125	$\\
$m_H$            	&$	6776	$&$	3558	$&$	3270	$\\
$m_A$            	&$	6732	$&$	3535	$&$	3249	$\\
$m_{H^{\pm}}$    	&$	6777	$&$	3559	$&$	3271	$\\
		  		  		  	
\hline		  		  		  	
$m_{\tilde{\chi}^0_{1,2}}$	&$	         456,         2478	$&$	         196,         1083	$&$	         153,          846	$\\

$m_{\tilde{\chi}^0_{3,4}}$	&$	        3665,         3667	$&$	        3434,         3434	$&$	        3303,         3304	$\\

$m_{\tilde{\chi}^{\pm}_{1,2}}$	&$	        2493,         3675	$&$	        1089,         3442	$&$	         853,         3310	$\\

$m_{\tilde{g}}$  	&$	4096	$&$	1922	$&$	1534	$\\
		  		  		  	
\hline $m_{ \tilde{u}_{L,R}}$	&$	        6172,         5945	$&$	        3364,         3285	$&$	        3031,         2980	$\\
                 		  		  		  	
$m_{\tilde{t}_{1,2}}$	&$	         488,         4325	$&$	         934,         2428	$&$	         650,         2122	 $\\
                 		  		  		  	
\hline $m_{ \tilde{d}_{L,R}}$	&$	        6173,         5944	$&$	        3365,         3284	$&$	        3032,         2979	$\\
                 		  		  		  	
$m_{\tilde{b}_{1,2}}$	&$	        4382,         5562	$&$	        2427,         3077	$&$	        2118,         2774	 $\\
                 		  		  		  	
\hline		  		  		  	
$m_{\tilde{\nu}_{1}}$	&$	5263	$&$	3026	$&$	2810	$\\
                 		  		  		  	
$m_{\tilde{\nu}_{3}}$	&$	5088	$&$	2936	$&$	2724	$\\
                 		  		  		  	
\hline		  		  		  	
$m_{ \tilde{e}_{L,R}}$	&$	        5254,         4965	$&$	        3024,         2922	$&$	        2808,         2741	 $\\
                		  		  		  	
$m_{\tilde{\tau}_{1,2}}$	&$	        5066,         4567	$&$	        2926,         2727	$&$	        2715,         2559	$\\
                		  		  		  	
\hline		  		  		  	
$\Delta(g-2)_{\mu}$  	&$	  1.12\times 10^{-11}	$&$	  2.52\times 10^{-11}	$&$	  2.90\times 10^{-11}	$\\

$\sigma_{SI}({\rm pb})$	&$	  3.12\times 10^{-13}	$&$	  5.38\times 10^{-13}	$&$	  4.89\times 10^{-13}	$\\

$\sigma_{SD}({\rm pb})$	&$	  3.67\times 10^{-10}	$&$	  7.93\times 10^{-12}	$&$	  7.05\times 10^{-13}	$\\

$\Omega_{CDM}h^{2}$	&$	  0.11	$&$	  22	$&$	  371	$\\
                		  		  		  	
\hline		  		  		  	
		  		  		  	
$R_{t b \tau}$     	&$	1.12	$&$	1.06	$&$	1.05	$\\

\hline
\hline
\end{tabular}
\caption{Benchmark points with good Yukawa unification. Point 1 has a small neutralino relic abundance with YU around 11\%.  For point 2, the relic abundance is relatively large and the stop is twice as heavy compared to point 1 while the gluino is lighter. For point 3, YU is around the best value we obtained in our analysis ($\sim$ 5\%).  Point 3 also has good YU with a lighter gluino and stop compared to point 2. The Higgs mass for the three points is within the favored range, $123 {\rm \ GeV} \le m_h \le 127 \rm \ GeV$. Stop is the NLSP for the three points.}
\label{tab1}
\end{table}

%%%%%%%%%%%%%%%%%%%%%%%%%%%%%%%%%%%%%%%%
%\clearpage

In the $m_{\tilde{t}_1} - M_{\chi_1^0}$ plane, we observe that {20\% or better} YU can be achieved with the neutralino-stop coannihilation scenario. {This} is a prediction of this model if we require neutralino to be the {sole} dark matter candidate.  A lower limit on the mass of the NLSP stop quark was obtained in refs. \cite{He:2011tp} in light of 7 TeV LHC data corresponding to $1 \ {\rm fb}^{-1}$ integrated luminosity. It was shown that the NLSP stop mass below 140 GeV is essentially excluded.

 Note that the analysis in refs.\cite{Gogoladze:2011aa} employ similar GUT scale boundary conditions for the SSB terms  but with a different relation for the Yukawa couplings. The model in  \cite{Gogoladze:2011aa} predicts neutralino-stau coannihilation scenario  to be consistent with good YU.
 %This serves as an important distinction between the two models. The different relation Yukawa couplings in the two analysis therefore favor different regions of the parameter space. Moreover, the sparticle spectroscopy is also different in the two cases.
 {It also predicts relatively low values of $m_A $ and $m_{\tilde{\tau}_1}$}, while the stop and the gluino masses are $\gtrsim$ 5 TeV.

Figure \ref{fig-144-2} shows the results in  the $R^{\prime}_{tb\tau} - m_h$ and  $R^{\prime}_{tb\tau} - m_{\tilde{\chi}^{\pm}}$ planes. The color coding is the same as in Figure \ref{fig:nugm1}.
 We observe that the model accommodates the Higgs mass range, $123 {\rm \ GeV} \le m_h \le 127 \rm \ GeV$, while exhibiting good ($\sim 5\%$) YU. {Requiring $\Omega h^2 \le 10$, the Higgs mass can still be within the favorable range with YU still at the $\sim 5\%$ level}. The $R^{\prime}_{tb\tau} - m_{\tilde{\chi}^{\pm}}$ plane indicates that the chargino,  which in this model is mostly a wino like particle, can be as light as 500 GeV. {Imposing the relic abundance  bound implies that the lightest chargino mass is $\sim$ 1 TeV}. We therefore conclude that {compatibility} of YU with neutralino dark matter scenario  in this model {predicts  that only the stop quark will be} accessible at the LHC. If we assume that neutralino is not the dark matter candidate we can see from Figure
\ref{fig:nugm3} that the gluino can be around 1 TeV, while the squarks can lie around 2 TeV or so.

In Table \ref{tab1} we present three  {characteristic} benchmark points which {summarize the salient features of this model}. The three points satisfy $123 {\rm \ GeV} \le m_h \le 127 \rm \ GeV$ {as well as} the sparticle mass and B-physics constraints described in section \ref{constraintsSection}. The mass of the gluino decreases {from 4 TeV (point 1) to 1.5 TeV (point 3)}. For point 1, YU is {at the level of} 11\% and the {neutralino} relic abundance satisfies the 5$\sigma$ WMAP bound. For point 2, the {neutralino} relic abundance is relatively large but YU is {at the few percent level}. Point 3 {shows acceptable} YU with a lighter gluino and stop compared to point 2. The stop is the NLSP for the three points with the lightest being $m_{\tilde{t}_1} = 488$ GeV for the first point.

{Note that in {some} SO(10) GUT models  with a unified Higgs sector, it
is possible to have relations among Yukawa couplings
\cite{Babu:2005gx,Babu:2006rp}  {different from} what we employed in Eq. (\ref{eq144-2}).
For instance, in ref. \cite{Babu:2005gx} {it is discussed} that there is
parameter space {available consistent with} the relation:}
\begin{align}
Y_t/48=Y_b=Y_{\tau},
\label{yu2}
\end{align}
 {which predicts $\tan\beta\approx 2$. We {find} that in this case the solutions
 yield YU no better than the $70\%$ level,
 which therefore indicates there is no YU at all.}

{ A {different scenario} \cite{Babu:2006rp} allows:
} \begin{align}
Y_t/8.35=Y_b=0.7\, Y_{\tau}.
\label{yu3}
\end{align}
{The predicted value of the parameter $\tan\beta$ for this case is $\approx 10$ and is therefore preferable.
However, in this case also the best unification is only at the $12\%$
level. Because of these unfavorable results we are not presenting
a more detailed analysis for {these relations (Eqs (\ref{yu2}) and (\ref{yu3}))}.}

\section{Conclusion \label{conclusions}}

We {have explored} a class of SO(10) GUT models  with a unified Higgs sector which yield the asymptotic relation $Y_b\approx Y_{\tau}\approx Y_t/6$ among the third generation quark and lepton Yukawa couplings. This relation among the Yukawa couplings {is compatible with the various phenomenological constraints only with} non-universal SSB mass terms at the GUT scale. The best YU ($\sim 5\%$) we found in our analysis is consistent with the lightest CP-even Higgs boson mass to be in the interval $123 {\rm \ GeV} \le m_h \le 127 \rm \ GeV$.
By scanning the {fundamental} parameter space of this model we showed that $\tan\beta$ is {constrained} in a very narrow interval, namely,  $\tan\beta \approx 14$.
Without imposing the constraint on the relic abundance of dark matter in these models,  the squark and  slepton masses, except for the stop, exceed 2 TeV while the gluino can be more than 1 TeV.
On the other hand, the LSP neutralino as a dark matter candidate in this model can only be realized through the neutralino-stop coannihilation scenario. We found that requiring good YU can lead to a light stop  ($\gtrsim 500$ GeV) with all other sfermions  having masses {possibly} beyond the {reach} of the {14 TeV} LHC.

%%%%%%%%%%%%%%%%%%%%%%%%%%%%%%%%%%%%%%%%%%%%%%
\section*{Acknowledgments}

{We are extremely grateful to Pran Nath for collaborating with us in the
early stages of this work and for numerous encouraging discussions.}

This work is supported in part by the DOE Grants No. DE-FG02-12ER41808 and by Rustaveli
National Science Foundation  No. 03/79 (I.G.). This work used the Extreme Science
and Engineering Discovery Environment (XSEDE), which is supported by the National Science
Foundation grant number OCI-1053575.

 I.G. wishes to thank the Center for Theoretical Underground Physics and Related
Areas (CETUP*) where some part of this project was done.

%%%%%%%%%%%%%%%%%%%%%%%%%%%%%%%%%


\begin{thebibliography}{99}
%%%%%%%%%%%%%%%%%%%%%%%%%%%%%%%%%




\bibitem{georgi}
H. Georgi, in \textit{Particles and Fields}, edited by C.E. Carlson, (A.I.P., New York,
1975); H. Fritzch and P. Minkowski, Ann. Phys. {\bf 93}(1975)193.

\bibitem{seesawI}
P.~Minkowski, Phys. Lett. B {\bf 67}, 421 (1977);
T.~Yanagida, in \emph{Proceedings of the Workshop on the Unified
  Theory and the Baryon Number in the Universe} (O.~Sawada and
  A.~Sugamoto, eds.), KEK, Tsukuba, Japan, 1979, p.~95;
M.~Gell-Mann, P.~Ramond, and R.~Slansky, \emph{Supergravity} (P.~van
  Nieuwenhuizen et al. eds.), North Holland, Amsterdam, 1979, p.~315;
S.~L. Glashow, \emph{The future of elementary particle physics}, in
  \emph{Proceedings of the 1979 Carg{\`e}se Summer Institute
 on Quarks and Leptons} (M.~L{\'e}vy et al. eds.),
 Plenum Press, New York, 1980, p.~687;
R.~N. Mohapatra and G.~Senjanovi{\'c},
 Phys. Rev. Lett. {\bf 44}, 912 (1980).





 %\cite{Fukugita:1986hr}
\bibitem{Fukugita:1986hr}
  M.~Fukugita and T.~Yanagida,
  %``Baryogenesis Without Grand Unification,''
  Phys.\ Lett.\ B {\bf 174} (1986) 45;
  For non-thermal leptogenesis in inflation, see
%  \bibitem{Lazarides:1991wu}
  G.~Lazarides, Q.~Shafi and ,
  %``Origin of matter in the inflationary cosmology,''
  Phys.\ Lett.\ B {\bf 258}, 305 (1991).
  %%CITATION = PHLTA,B258,305;%%


\bibitem{big-422}
B. Ananthanarayan, G. Lazarides and Q. Shafi, Phys. Rev. D {\bf 44},
1613 (1991; Phys. Lett. B {\bf 300}, 24 (1993)5; Q.~Shafi and
B.~Ananthanarayan, Trieste HEP Cosmol.1991:233-244.



  %\cite{Chamseddine:1982jx}
\bibitem{Chamseddine:1982jx}
 A.~Chamseddine, R.~Arnowitt and P.~Nath, Phys.\ Rev.\ Lett.\ {\bf 49} (1982) 970;
R.~Barbieri, S.~Ferrara and C.~Savoy, Phys.\ Lett.\ {\bf B119}
(1982) 343; N.~Ohta, Prog.\ Theor.\ Phys.\ {\bf 70} (1983) 542;
L.~J.~Hall, J.~D.~Lykken and S.~Weinberg, Phys.\ Rev.\ {\bf D27}
(1983) 2359.




\bibitem{bigger-422}For an incomplete list of resent references see,
 H.~Baer, S.~Kraml and S.~Sekmen, JHEP {\bf 0909}, 005 (2009);
 I.~Gogoladze, R.~Khalid and Q.~Shafi,
  %``Yukawa Unification and Neutralino Dark Matter in SU(4)(c) x SU(2)(L) x SU(2)(R),''
  Phys.\ Rev.\ D {\bf 79}, 115004 (2009);
%  [arXiv:0903.5204 [hep-ph]].
S.~Antusch and M.~Spinrath,
Phys.\ Rev.\  D {\bf 79}, 095004 (2009);
%D.~Guadagnoli, S.~Raby and D.~M.~Straub, JHEP {\bf 0910}, 059 (2009);
I.~Gogoladze, R.~Khalid, S.~Raza and Q.~Shafi,
  %``t - b - tau Yukawa unification for mu < 0 with a sub-TeV sparticle spectrum,''
  JHEP {\bf 1012}, 055 (2010);
K.~Choi, D.~Guadagnoli, S.~H.~Im and C.~B.~Park,
  JHEP {\bf 1010}, 025 (2010);
%arXiv:1005.0618 [hep-ph].
 %\cite{arXiv:1105.5122}
%\bibitem{arXiv:1107.1228}
  I.~Gogoladze, Q.~Shafi and C.~S.~Un,
  %``SO(10) Yukawa Unification with mu < 0,''
  Phys.\ Lett.\ B\ {\bf 704}, 201  (2011);
 % [arXiv:1107.1228 [hep-ph]].
  %%CITATION = PHLTA,B704,201;
   M.~Badziak, M.~Olechowski and S.~Pokorski,
  %``Yukawa unification in SO(10) with light sparticle spectrum,''
  JHEP\ {\bf 1108}, 147  (2011);
  %[arXiv:1107.2764 [hep-ph]].
%\cite{arXiv:1111.6547}
%\bibitem{arXiv:1111.6547}
  S.~Antusch, L.~Calibbi, V.~Maurer, M.~Monaco and M.~Spinrath,
  %``Naturalness and GUT Scale Yukawa Coupling Ratios in the CMSSM,''
%  arXiv:1111.6547 [hep-ph];
   Phys.\ Rev.\ D {\bf 85}, 035025 (2012).
  %%CITATION = ARXIV:1111.6547;%%
%\cite{arXiv:1111.3639}
%\bibitem{arXiv:1111.3639}
  J.~S.~Gainer, R.~Huo and C.~E.~M.~Wagner,
  %``An Alternative Yukawa Unified SUSY Scenario,''
  %arXiv:1111.3639 [hep-ph].
    JHEP {\bf 1203}, 097 (2012);
     H.~Baer, S.~Raza and Q.~Shafi,
  %``A Heavier gluino from $t-b-\tau$ Yukawa-unified SUSY,''
  Phys.\ Lett.\ B {\bf 712}, 250 (2012);
%  arXiv:1201.5668 [hep-ph];
  %%CITATION = ARXIV:1111.3639;%
   M.~Badziak,
  %``Yukawa unification in SUSY SO(10) in light of the LHC Higgs data,''
  Mod.\ Phys.\ Lett.\ A {\bf 27}, 1230020 (2012);
  %\cite{Gogoladze:2012ii}
%\bibitem{Gogoladze:2012ii}
  I.~Gogoladze, Q.~Shafi, C.~S.~Un and ,
  %``125 GeV Higgs Boson from t-b-tau Yukawa Unification,''
  JHEP {\bf 1207}, 055 (2012);
 % [arXiv:1203.6082 [hep-ph]].
  %%CITATION = ARXIV:1203.6082;%%
  %8 citations counted in INSPIRE as of 25 Mar 2013
%\cite{Elor:2012ig}
%\bibitem{Elor:2012ig}
  G.~Elor, L.~J.~Hall, D.~Pinner and  J.~T.~Ruderman,
  %``Yukawa Unification and the Superpartner Mass Scale,''
  JHEP {\bf 1210}, 111 (2012);
%  [arXiv:1206.5301 [hep-ph]].
  %%CITATION = ARXIV:1206.5301;%%
  %5 citations counted in INSPIRE as of 25 Mar 2013
%  \bibitem{Anandakrishnan:2012tj}
  A.~Anandakrishnan, S.~Raby and A.~Wingerter,
  %``Yukawa Unification Predictions for the LHC,''
  Phys.\ Rev.\ D {\bf 87}, 055005 (2013);
  %\cite{Karagiannakis:2012sv}
%\bibitem{Karagiannakis:2012sv}
  N.~Karagiannakis, G.~Lazarides and C.~Pallis,
  %``Constrained Minimal Supersymmetric Standard Model with Generalized Yukawa Quasi-Unification,''
  Phys.\  Rev.\  D 87, {\bf 055001} (2013)
 % [Phys.\ Rev.\ D {\bf 87}, 055001 (2013)]
 % [arXiv:1212.0517 [hep-ph]].
  %%CITATION = ARXIV:1212.0517;%%
  %1 citations counted in INSPIRE as of 24 May 2013
  %[arXiv:1212.0542 [hep-ph]].
 A.~Anandakrishnan and S.~Raby,
  %``Yukawa Unification Predictions with effective "Mirage" Mediation,''
  arXiv:1303.5125 [hep-ph].






%\cite{Gogoladze:2011aa}
\bibitem{Gogoladze:2011aa}
  I.~Gogoladze, Q.~Shafi and C.~S.~Un,
  %``Higgs Boson Mass from t-b-$\tau$ Yukawa Unification,''
  JHEP {\bf 1208}, 028 (2012)
%  [arXiv:1112.2206 [hep-ph]];
  %%CITATION = ARXIV:1112.2206;%%
  %31 citations counted in INSPIRE as of 19 May 2013
%\cite{Ajaib:2013zha}
%\bibitem{Ajaib:2013zha}
  M.~Adeel Ajaib, I.~Gogoladze, Q.~Shafi and C.~S.~Un,
  %``A Predictive Yukawa Unified SO(10) Model: Higgs and Sparticle Masses,''
  arXiv:1303.6964 [hep-ph].
  %%CITATION = ARXIV:1303.6964;%%
  %1 citations counted in INSPIRE as of 19 May 2013


%\cite{Babu:2005gx}
\bibitem{Babu:2005gx}
  K.~S.~Babu, I.~Gogoladze, P.~Nath and R.~M.~Syed,
  %``A Unified framework for symmetry breaking in SO(10),''
  Phys.\ Rev.\ D {\bf 72} (2005) 095011.
 % [hep-ph/0506312].
  %%CITATION = HEP-PH/0506312;%%
  %30 citations counted in INSPIRE as of 27 May 2013


%\cite{Babu:2006rp}
\bibitem{Babu:2006rp}
  K.~S.~Babu, I.~Gogoladze, P.~Nath and R.~M.~Syed,
  %``Fermion Mass Generation in SO(10) with a Unified Higgs Sector,''
  Phys.\ Rev.\ D {\bf 74}, 075004 (2006);
 % [hep-ph/0607244].
  %%CITATION = HEP-PH/0607244;%%
  %24 citations counted in INSPIRE as of 27 May 2013
%\cite{Nath:2007eg}
%\bibitem{Nath:2007eg}
  P.~Nath and R.~M.~Syed,
  %``Suppression of Higgsino mediated proton decay by cancellations in GUTs and strings,''
  Phys.\ Rev.\ D {\bf 77}, 015015 (2008);
 % [arXiv:0707.1332 [hep-ph]];
  %%CITATION = ARXIV:0707.1332;%%
  %14 citations counted in INSPIRE as of 11 Jul 2013
%\cite{Wu:2009zzh}
%\bibitem{Wu:2009zzh}
  Y.~Wu and D.~X.~Zhang,
  %``Proton decay and fermion masses in supersymmetric SO (10) model with unified Higgs sector,''
  Phys.\ Rev.\ D {\bf 80}, 035022 (2009);
 % [arXiv:0909.1179 [hep-ph]].
  %%CITATION = ARXIV:0909.1179;%%
  %3 citations counted in INSPIRE as of 11 Jul 2013
%\cite{Nath:2009nf}
%\bibitem{Nath:2009nf}
  P.~Nath and R.~M. Syed,
  %``Yukawa Couplings and Quark and Lepton Masses in an SO(10) Model with a Unified Higgs Sector,''
  Phys.\ Rev.\ D {\bf 81}, 037701 (2010).
% [arXiv:0909.2380 [hep-ph]].
  %%CITATION = ARXIV:0909.2380;%%
  %13 citations counted in INSPIRE as of 11 Jul 2013


%\cite{Babu:2011tw}
\bibitem{Babu:2011tw}
  K.~S.~Babu, I.~Gogoladze, P.~Nath and R.~M.~Syed,
  %``Variety of SO(10) GUTs with Natural Doublet-Triplet Splitting via the Missing Partner Mechanism,''
  Phys.\ Rev.\ D {\bf 85}, 075002 (2012).
%  [arXiv:1112.5387 [hep-ph]].
  %%CITATION = ARXIV:1112.5387;%%
  %8 citations counted in INSPIRE as of 27 May 2013



%\cite{Martin:2009ad}
\bibitem{Martin:2009ad}
 B.~Ananthanarayan, P.~N.~Pandita,
  %``Sparticle Mass Spectrum in Grand Unified Theories,''
  Int.\ J.\ Mod.\ Phys.\  {\bf A22}, 3229-3259 (2007);
%  [arXiv:0706.2560 [hep-ph]].
%\cite{Bhattacharya:2007dr}
%\bibitem{Bhattacharya:2007dr}
  S.~Bhattacharya, A.~Datta and B.~Mukhopadhyaya,
  %``Non-universal gaugino masses: A Signal-based analysis for the Large Hadron
  %Collider,''
  JHEP {\bf 0710}, 080 (2007);
  %[arXiv:0708.2427 [hep-ph]].
 S.~P.~Martin,
  %``Non-universal gaugino masses from non-singlet F-terms in non-minimal unified models,''
  Phys.\ Rev.\  {\bf D79}, 095019 (2009);
    U.~Chattopadhyay, D.~Das and D.~P.~Roy,
  %``Mixed Neutralino Dark Matter in Nonuniversal Gaugino Mass Models,''
  Phys.\ Rev.\  D {\bf 79}, 095013 (2009);
  %[arXiv:0902.4568 [hep-ph]].
  %%CITATION = PHRVA,D79,095013;%%
  %%CITATION = JHEPA,0710,080;%%
%\cite{Corsetti:2000yq}
%\bibitem{Corsetti:2000yq}
  A.~Corsetti and P.~Nath,
  %``Gaugino Mass Nonuniversality and Dark Matter in SUGRA, Strings and D
  %Brane
  %Models,''
  Phys.\ Rev.\  D {\bf 64}, 125010 (2001)
  %[arXiv:hep-ph/0003186].
  %%CITATION = PHRVA,D64,125010;%%
   and references therein.
 % [arXiv:0903.3568 [hep-ph]].



\bibitem{ISAJET}
%\cite{hep-ph/0312045}
%\bibitem{hep-ph/0312045}
  F.~E.~Paige, S.~D.~Protopopescu, H.~Baer and X.~Tata,
  %``ISAJET 7.69: A Monte Carlo event generator for pp, anti-p p, and e+e- reactions,''
  hep-ph/0312045.
  %%CITATION = HEP-PH/0312045;%%



\bibitem{Hisano:1992jj}
J.~Hisano, H.~Murayama  , and T.~Yanagida,
%{\it Nucleon decay in the minimal
  %supersymmetric SU(5) grand unification},
  { Nucl. Phys.} {\bf B402} (1993) 46.
%  [\href{http://xxx.lanl.gov/abs/hep-ph/9207279}\left\{ \tt
 % hep-ph/9207279}}].
%\bibitem{Yamada:1992kv}
Y.~Yamada,
%{\it SUSY and GUT threshold effects in SUSY SU(5) models},
{ Z. Phys.} {\bf C60} (1993) 83;
 J.~L.~Chkareuli and I.~G.~Gogoladze,
  %``Unification picture in minimal supersymmetric SU(5) model with string
  %remnants,''
  Phys.\ Rev.\  D {\bf 58}, 055011 (1998).
%  [arXiv:hep-ph/9803335].





\bibitem{Pierce:1996zz}
D.~M. Pierce, J.~A. Bagger, K.~T. Matchev, and R.-j. Zhang,
% {\it Precision   corrections in the minimal supersymmetric standard
% model},
  { Nucl. Phys.} {\bf B491} (1997) 3.
  %[\href{http://xxx.lanl.gov/abs/hep-ph/9606211}\left\{ \tt
  %hep-ph/9606211}}].

  %\cite{Nakamura:2010zzi}
\bibitem{Nakamura:2010zzi}
  K. Nakamura {\it et al.} [ Particle Data Group Collaboration ],
  %``Review of particle physics,''
  J.\ Phys.\ G {\bf G37}, 075021 (2010).

   %\cite{Belanger:2008sj}
\bibitem{Belanger:2008sj}
  G.~Belanger, F.~Boudjema, A.~Pukhov and A.~Semenov,
  %``Dark matter direct detection rate in a generic model with micrOMEGAs 2.2,''
  Comput.\ Phys.\ Commun.\  {\bf 180}, 747 (2009).
  %[arXiv:0803.2360 [hep-ph]].
  %%CITATION = ARXIV:0803.2360;%%
  %291 citations counted in INSPIRE as of 14 Mar 2013

%\cite{:1900yx}
\bibitem{:1900yx}
 [Tevatron Electroweak Working Group and CDF Collaboration and D0 Collab],
  %``Combination of CDF and D0 Results on the Mass of the Top Quark,''
  arXiv:0903.2503 [hep-ex].
  %%CITATION = ARXIV:0903.2503;%%






%\cite{Gogoladze:2011db}
\bibitem{Gogoladze:2011db}
  I.~Gogoladze, R.~Khalid, S.~Raza and Q.~Shafi,
  %``Higgs and Sparticle Spectroscopy with Gauge-Yukawa Unification,''
  JHEP {\bf 1106} (2011) 117.
  %[arXiv:1102.0013 [hep-ph]].
  %%CITATION = JHEPA,1106,117;%%



%\cite{Belanger:2009ti}
\bibitem{Belanger:2009ti}
  G.~Belanger, F.~Boudjema, A.~Pukhov and R.~K.~Singh,
  %``Constraining the MSSM with universal gaugino masses and implication for
  %searches at the LHC,''
  JHEP {\bf 0911}, 026 (2009);
  %[arXiv:0906.5048 [hep-ph]].
  %%CITATION = JHEPA,0911,026;%%
H.~Baer, S.~Kraml, S.~Sekmen and H.~Summy,
  %``Dark matter allowed scenarios for Yukawa-unified SO(10) SUSY GUTs,''
  JHEP {\bf 0803}, 056 (2008).
  %[arXiv:0801.1831 [hep-ph]].
  %%CITATION = JHEPA,0803,056;%%

\bibitem{Baer:2002fv}
H.~Baer, C.~Balazs, and A.~Belyaev,
%{\it Neutralino relic density in minimal
  %supergravity with co-annihilations},
   { JHEP} {\bf 03} (2002) 042;
 %  [\href{http://xxx.lanl.gov/abs/hep-ph/0202076}\left\{ \tt hep-ph/0202076}}].
%\bibitem{Baer:2001kn}
 H.~Baer, C.~Balazs, J.~Ferrandis, and X.~Tata
%, {\it Impact of muon anomalous
  % magnetic moment on supersymmetric models},
  { Phys. Rev.} {\bf D64} (2001)  035004.
  % [\href{http://xxx.lanl.gov/abs/hep-ph/0103280}\left\{ \tt hep-ph/0103280}}].





%\cite{Leva}
\bibitem{Leva}
J.L. Leva,
%A fast normal random number generator, ACM Trans.
 Math. Softw. 18 (1992) 449;
J.L. Leva,
%Algorithm 712. A normal random number generator, ACM Trans.
Math. Softw. 18 (1992) 454.





%\cite{:2007kv}
\bibitem{:2007kv}
  T.~Aaltonen {\it et al.}  [CDF Collaboration],
  %``Search for $B_s \to \mu^+\mu^-$ and $B_d \to \mu^+\mu^-$ Decays with
  %2fb$^{-1}$ of $p\bar{p}$ Collisions,''
  Phys.\ Rev.\ Lett.\  {\bf 100}, 101802 (2008).
 % [arXiv:0712.1708 [hep-ex]].

%\bibitem{Barberio:2007cr}
%  E.~Barberio {\it et al.}  [Heavy Flavor Averaging Group (HFAG)
%                  Collaboration],
  %``Averages of b-hadron properties at the end of 2006,''
%  arXiv:0704.3575 [hep-ex].

%\cite{Barberio:2008fa}
\bibitem{Barberio:2008fa}
  E.~Barberio {\it et al.}  [Heavy Flavor Averaging Group],
  %``Averages of $b-$hadron and $c-$hadron Properties at the End of 2007,''
  arXiv:0808.1297 [hep-ex].
  %%CITATION = ARXIV:0808.1297;%%








%\cite{Bennett:2006fi}
\bibitem{Bennett:2006fi}
  G.~W.~Bennett {\it et al.}  [Muon G-2 Collaboration],
  %``Final report of the muon E821 anomalous magnetic moment measurement at
  %BNL,''
  Phys.\ Rev.\  D {\bf 73}, 072003 (2006).





%\cite{Baer:2006te}
\bibitem{Baer:2006te}
%\cite{ArkaniHamed:2006mb}
%\bibitem{ArkaniHamed:2006mb}
  N.~Arkani-Hamed, A.~Delgado and G.~F.~Giudice,
  %``The Well-tempered neutralino,''
  Nucl.\ Phys.\ B {\bf 741}, 108 (2006);
 % [hep-ph/0601041].
  %%CITATION = HEP-PH/0601041;%%
  H.~Baer, A.~Mustafayev, E.~-K.~Park and X.~Tata,
  %``Target dark matter detection rates in models with a well-tempered neutralino,''
  JCAP {\bf 0701}, 017 (2007).
%  [hep-ph/0611387].
  %%CITATION = HEP-PH/0611387;%%

%\cite{He:2011tp}
\bibitem{He:2011tp}
  B.~He, T.~Li and Q.~Shafi,
  %``Impact of LHC Searches on NLSP Top Squark and Gluino Mass,''
  JHEP {\bf 1205}, 148 (2012);
  %[arXiv:1112.4461 [hep-ph]];
  %%CITATION = ARXIV:1112.4461;%%
  %22 citations counted in INSPIRE as of 28 May 2013
%
%\cite{Ajaib:2011hs}
%\bibitem{Ajaib:2011hs}
  M.~A.~Ajaib, T.~Li and Q.~Shafi,
  %``Stop-Neutralino Coannihilation in the Light of LHC,''
  Phys.\ Rev.\ D {\bf 85}, 055021 (2012).
  %[arXiv:1111.4467 [hep-ph]].
  %%CITATION = ARXIV:1111.4467;%%
  %22 citations counted in INSPIRE as of 28 May 2013

\end{thebibliography}
\end{document}